# Community Detection in Social Network using Temporal Data

Victor Stany Rozario, AZM Ehtesham Chowdhury, Muhammad Sarwar Jahan Morshed

*Abstract*— Social networks facilitate the social space where actors or the users have ties among them. The ties and their patterns are based on their life styles and communication. Similarly, in online social media networks like Facebook, Twitter, Myspace etc., multiple users belong to multiple specific communities. The social network represents large real-world directed social graphs. Detection of communities or clusters from these graphs is a problem of considerable research interest. The communities are formed using the neighboring nodes that have common edges and common attributes. Most of the existing community detection algorithms usually consider node contents to analyze the attributes of community. Some algorithms use the links between the nodes to determine the dense regions in the graph. But utilizing both the edge content and the vertex content to detect community are yet to be considered and verified, since the traditional extraction methods of vertex and edge data do not consider the connectivity among the nodes. This paper presents an Interlinked Spatial Clustering Model (ILSCM) which provides relevant content selection and extraction of the temporal topics for identifying the betweenness among the nodes based on the context keys to detect community.

## I. INTRODUCTION

Social networks are common paradigm to interconnect people from diverse regions to share their thoughts, believes and life styles. Capture those emotions, thoughts, actions and interests are valuable assets to analyze the impact of social network on the border-less region. Individual user generates these assets. These are named as User-generated contents (UGC). These contents can be divided into two categories based on the events and user's field of interests. First one is temporal topics (e.g., breaking events like Independence Day) while second one is stable topics (e.g., user interests) [1]. Some example of these two topics have been stated in Table I. User's regular interest, communication through messenger, daily routine discussions are stable data. These stable topics expand and adapt slowly to other users on the network graph. Temporal topics are breaking events in the region and world. It can be popular event discussions and wide diffusion where social networks moreover boost the discussion and diffusion [2].

Different topic models and burst feature clustering have also been applied to detect topics with temporal features. The topic model without considering individual burst word can lead to the problem with high occurrence rate of these words. It is mostly denoting semantically abstract concepts, are likely to appear at top positions in summaries. Moreover, to the best of our knowledge, there is no previous proposal that defines the threshold value for the time stamp which can distinguish the temporal topic and the stable topic. In the social network, community can be detected based on the contents from the communication graph. The mining algorithms use global criteria to identify a subgraph [3] as a community. But using the fixed criteria means a fixed social indicator of the community, [4] while detection communities tends to be overlapped. Overlapping communities can be depicted as overlapping clusters [5] on complex social network. The detection of the clusters based on the betweenness are the centrality measure [6]. As a result, selection of those criteria must face different tradeoffs. The criteria are enhancing and expanding with multiple dimensions for dynamically growing networks. Edge contents on social network graph present

**Victor Stany Rozario**
Dept. of CS, FSIT
American International University-Bangladesh (AIUB).
Dhaka, Bangladesh
e-mail: stany@aiub.edu

**AZM Ehtesham Chowdhury**
Dept. of CS, FSIT
American International University-Bangladesh (AIUB).
Dhaka, Bangladesh
e-mail: ehtesham@aiub.edu

**Muhammad Sarwar Jahan Morshed**
Dept. of CS, FSIT
American International University-Bangladesh (AIUB).
Dhaka, Bangladesh
e-mail: sarwar.morshed@aiub.edu

TABLE I. Temporal Topics vs. Stable Topics

| Temporal Topics | Stable Topics |
|---|---|
| Independence Day | Adopt Pet |
| Birthday | Visiting Place |
| Religious Day | Rescue |
| Strike Day | Watching Movie |
| Holiday | Eating Foods |
| Victory Day | Help |

great insights for identifying different communities. These contents characterize adequately the participants' interaction in the community. Edge contents leverage well flexibility for detection of communities from social network graphs [7]. In social network graph, edges can be defined as *likes*, shared images and videos, user *tags*, *comments*, *groups*, *liked pages* and so on. Vertex contents are those information like *name*, *gender*, *hometown*, *school* etc. that identify a user. Access to the social network is closely in nature. Therefore, these social networks provide APIs to access the edge and vertex contents of the social graph. For instance, social network like Facebook has Graph API to go through the social graph of Facebook. Using this API, we can easily look up the node and edge contents. In the following Listing 1, there is a JSON script showing the parts of the contents for two nodes or user:
Listing 1: "Parts of The Vertex Content in Facebook Graph for a user"

```
"id": "5843215622",
"name" : "Jibon Kumar",
"first name" : "Jibon",
"last name" : "Kumar",
"link" :
    "https://www.facebook.com/jibon.kumar",
"username" : "Hellrider Jibon",
"gender" : "male",
"type" : "user",
"locale" : "en_US",
"hometown" : "Rangpur, Bangladesh",
"email" : "jibon444@ymail.com"
```

Above Listing 1 shows two Facebook users are not friends and familiar to each other. From the only vertex information, it is possible to cluster them on a community like same home town. But it is not possible to conclude that they have same field of interests. If they have *like* or *share* on same status as *Today's Victory Day march was inspiring*, then an edge will be created between the two users on the graph. Now, it is possible to cluster them on a same community who have interest on *National Events*. So, by analyzing the vertex content and the edge content of users along with the temporal data, it is possible to identify more communities in a social network.

In section II, the related works have been discussed. In section III, we have shown how our proposed model works to detect and to form clusters to identify communities by analyzing the vertex content and the edge content using temporal data. And in section IV, we have given direction of future aspects of our proposed model and concluded the paper.

## II. RELATED WORKS

Different topic models are proposed to distinguish stable data and temporal data. Yin and Bin cui proposed Unified user-temporal mixture model [1] to identify the stable data and temporal data from social media. To boost interesting temporal topics, authors have proposed a smoothing technique that merges correlated words into one temporal topic. Zhiting Hu, Junjie Yao and Bin Cui proposed GrosToT (Group Specific Topics-overTime), a unified probabilistic model that infer latent user groups and temporal topics simultaneously [8]. It modeled group-specific temporal topic variation from social content. Zhiting Hu and Wang have proposed community level topic dynamic extraction. CoSTot (Community Specific Topics-over-Time) was the probabilistic method which is proposed to extract the concealed topics and communities [2]. This method simultaneously inspected text, time, and network information to uncover the interactions between topic and community. Jure and Yang selected thirteen structural definitions of network communities. Their model examined the sensitivity, robustness and performance in identifying the ground-truth to characterize and decide network communities [9]. Different algorithms were proposed for detecting the overlapping communities. Steve Gregory proposed an algorithm which detect the overlapping communities on social network [5] by extending Newman and Grivan's [6] hierarchical clustering algorithm which was based on the betweenness centrality measure for vertex nodes or users. Wanyun Cui and Yanghua Xiao proposed an approach that takes a query vertex by extracting community criterion. Then the community criterion mediatesd to detect meaningful overlapping communities [10]. Lescovec and Yang presented a paradigm for uncovering the modular structure [11] of complex networks for defining overlapping communities. Zhiting Hu and Junjie Yao introduced COLD (Community Level Diffusion) [12] to find the temporal diffusion. It model topics and communities in a unified latent framework, and excerpt inter-community influence dynamics. Yang and Jure Lescovec established a nonnegative matrix factorization approach [13] for Cluster Affiliation Model for Big Networks named BIGCLAM to detect an overlapping community which can be very large networks of millions of nodes and edges. As social network is represented as a large complex graph, analysis for this complex graph is plays a vital role for the detection of communities. Each community represents as cluster in this social media graph. Goldstein Doron established fractional LP solution [3] to detect a DamkS [14] solution to very dense subgraph problem. David Hallac and Jure Leskovec proposed Network Lasso [15] to cluster and optimize in large Graphs. Jure and Yang proposed a Community-Affiliation Graph Model [16] which was a model based community detection approach that builds on bipartite node-community affiliation networks. Vertex and edge content analysis are playing great role on classification of network community clusters. For community detection Julian McAuley, Jure and Yang developed CESNA [17]. It statistically modeled the interaction between then network structure and the node attributes. Multiplicative Attribute Graph (MAG) [18] was used by Kim and Jure for this purpose. For effective and efficient community detection, analysis of edge contents [7] of social network graph plays an important role. But in graph there are unobserved parts (missing edge and node). Kim and Jure have combined the EM with the Kronecker graphed model and design a scalable Metropolized Gibbs sampling approach to estimate model parameters to infer missing nodes and edges of the network.

Existing algorithms or systems did not define the threshold value for the time stamp which could distinguish the temporal topic and the stable topic. We have established a ILSCM (Interlinked Spatial Clustering Model) to detect and classify communities by distinguishing stable data and temporal data between nodes, Community overlapping, and efficient selection of edge data and vertex data. User community for a specific time and specific event can be detected using this model.

### III. PROPOSED MODEL

The proposed model's aim is to detect and to define different communities from social network graph by establishing a spatial clustering model [19] which means our model cluster communities based on their regions as well as the place of interests. Our proposed methodology contains several phases for differentiating and extracting temporal topics among those regions or spatial clusters based on user supplied context key. After that measuring the proposed model redefines and considers edge contents for betweenness among the vertices. Then weight on adjacent edges will be applied for identifying close betweenness among the contents. And a threshold level is established for different clusters of nodes and edges. Thus our model detects and creates cluster of different community. Our proposed model has following properties:

• Spatial Clustering model

• User supplied context key based temporal topic selection

• Consider edge contents for identifying betweenness among the vertices

• Apply weight on adjacent edges for identifying close betweenness

A set of users supplied context keys $C_k$ will be taken as input for searching temporal topics Based on these keys, a search function $f_s(C_k)$ identifies the set of burst words $B_{xy}$ from the edges on a temporal topic between two vertices $v_x$ & $v_y$. Number of burst words are considered as weight $w_{xy}$ for the edge between two vertices $v_x$ & $v_y$. Apply weight $w_{xy}$ on adjacent edges for identifying betweenness using adjacency matrix. A threshold value $\lambda$ is used to compare the betweenness for considering a vertex in a community. We assume, a social network is a graph $S_G = (V, E)$, where V is a set of nodes and E is a set of edges between nodes in V. Each node in V corresponds to a user, and an edge $E = (v_x, v_y)$. E stands for the social relationship between two users $v_x$ & $v_y$. The strength of the tie between users $v_x$ & $v_y$ is defined as a normalized non-negative value $\pi(v_x, v_y)$. Our proposed approach to find burst clusters among the edge's node $v_x$ & $v_y$ inspired by Wang and Zhai's Mining Correlated Bursty Topic Patterns [20]. We have provided set of context keys $C_k = \{C_1, C_2, C_3 \ldots .. C_k\}$. Each context key depicts the contexts of the particular event. For each context key, the searching function $fs$ returns a set of bursts $B_{xy}$ from the both $v_x$ & $v_y$. $B_{xy} = \{B_1, B_2, B_3 \ldots \ldots B_n\}$ Here n is the number of bursts for that particular edge. The data actually are depicted from the users and temporal features. after that the Pearson correlation coefficient r(x, y) is computed between two words x and y from the following equation :

$$r(x,y) = \frac{\sum_{i=1}^{n} x_i y_i \frac{1}{n} \sum_{i=1}^{n} x_i \sum_{i=1}^{n} y_i}{\sqrt{(\sum_{i=1}^{n} x_i^2 - \frac{1}{n}(\sum_{1=1}^{n} x_i)^2)(\sum_{i=1}^{n} y_i^2 - \frac{1}{n}(\sum_{i=1}^{n} y_i)^2)}}$$

where $x_i$ & $y_i$ are $i^{th}$ entries in x and ys frequency distribution vectors defined above. From this correlation, we have got the level of key matches to texts on different nodes. Then following Zhai and Wang's approach [20], we have got the set of burst words $B_{xy}$. Now we have count the number of elements on set $B_{xy}$. It's considered as $w_{xy}$. We have repeat the process for every edge on graph G. Then for every set of burst words, we have got set of weights W for each of the edges. An adjacency matrix is shown on figure 2 for visualizing the weights on nodes. From the adjacency matrix, we can calculate the betweenness among the nodes. A threshold $\lambda$ have been used to compare the betweenness for considering a vertex in a community. Here is ILSCM have shown as block diagram on Figure 1.

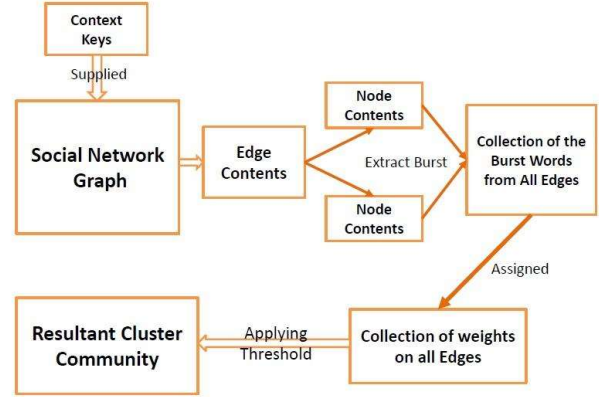

Fig. 1. **Interlinked Spatial Clustering Model**

---

**Algorithm:** Detect Community from Social Media Graph

Let, A connected social graph, $SG = (V; E)$. V is the set of Vertices/Nodes and E is the set of Edges.

**Input:** Set of Context keys $C_k$

**Output:** A resultant subgraph, $RG = (V_r, E_r)$. $V_r$ is set of Vertices/Nodes and $E_r$ is the set of Edges for resultant subgraph.

**Step 1:** An element from set $Ck$ regarding determined topic, a search function $f_s$ will be called.
**Step 1.1:** For all edges on set $E$ select the adjacent nodes $V_x$ & $V_y$.
**Step 1.2:** Search a set of Burst Words $B_{xy}$.
**Step 1.3:** Count C, the numbers Burst Words in $B_{xy}$.
**Step 1.4:** C will be assigned as weight $W$.
**Step 2:** Add $W$ on set $W_{xy}$.
**Step 3:** Select the edges whose $W$ is bigger than threshold $\lambda$.

**Step 4:** Make a new graph $R_G = (V_r, E_r)$, where selected edges will be adding on set $E_r$. And nodes are in set $V_r$.

---

**Comment**: New graph, $R_G = (V_r, E_r)$ is desired community graph. Here in Figure 2, users are represented by the nodes and their links (e.g. comment, share, like etc.) are represented by edges in Graph G. Weights are assigned on the edges. $\lambda_{10}$ will made a cluster for the specific context stated earlier that represents the detected community on subgraph $R_G$.

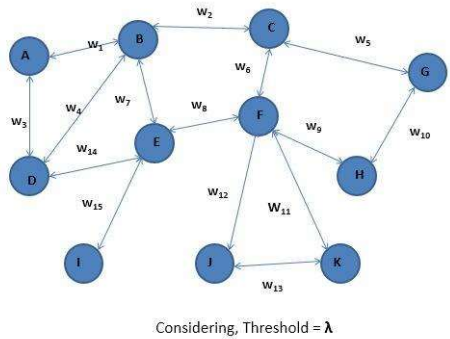

Fig. 2. **G: A Small Social Network Graph**

IV. RESULT ANALYSIS

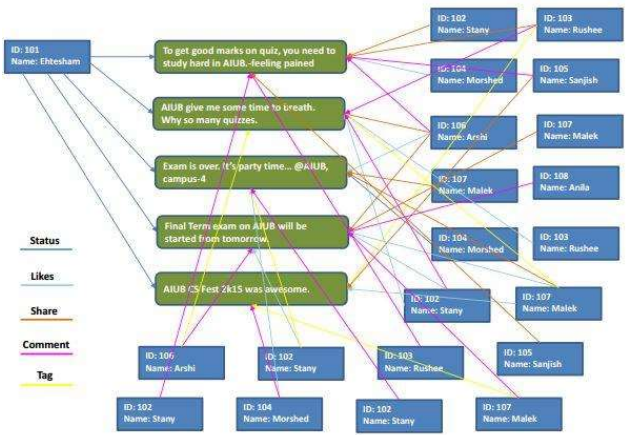

Fig. 3. **User to User (nodes) Connecting Edges**

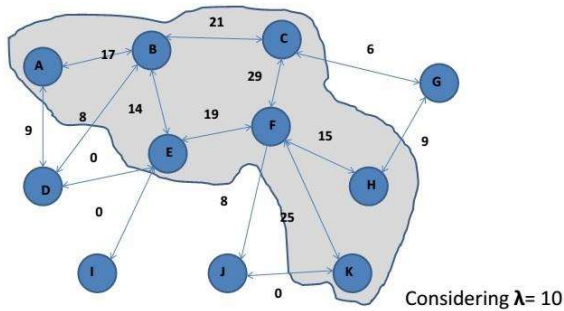

Fig. 4. **A Small Social Network Graph G**

From the Figure 3, we have shown a part of analyzed graph where node id '101' named 'Ehtesham' has five statuses. The connecting line shows that how other nodes or users connected with the user by edges (likes, comments, tags, shares). Each components are depicted as edges among the users to users. The edges are considered as weights considering the context key 'AIUB' or 'Exam' or 'Quiz'. Each of the lines one user to another user is considered as weight. Following the basic which is stated above, we have calculated all weights from one node to another node. The weights are facilitated by the edges among themselves. The part of analyzed social graph in Figure 4. Here in Figure 4, users are represented by the nodes (A, B, C, D etc.) and their links (e.g. comment, share, like etc.) are represented by edges in Graph. Weight values are assigned on the edges. $\lambda_{10}$ will made a cluster for the specific context stated earlier that represents the detected community on subgraph $R_G$. Now here each user is connected with other users by their edges and connectivity of them also calculated by the weights of their connectivity.

|   | A | B | C | D | E | F | G | H | I | J | K |
|---|---|---|---|---|---|---|---|---|---|---|---|
| A |   | 17 |   | 9 |   |   |   |   |   |   |   |
| B | 17 |   | 21 | 8 | 14 |   |   |   |   |   |   |
| C |   | 21 |   |   |   | 29 | 6 |   |   |   |   |
| D | 9 | 8 |   |   | 0 |   |   |   |   |   |   |
| E |   | 14 |   | 0 |   | 19 |   |   | 0 |   |   |
| F |   |   | 29 |   | 19 |   |   | 15 |   | 8 | 25 |
| G |   |   | 6 |   |   |   |   | 9 |   |   |   |
| H |   |   |   |   |   |   |   |   |   |   |   |
| I |   |   |   |   | 0 |   |   |   |   |   |   |
| J |   |   |   |   |   | 8 |   |   |   |   | 0 |
| K |   |   |   |   |   | 0 |   |   |   | 25 |   |

Fig. 5. **Adjacency Matrix Representing Graph G**

From the matrix on Figure 5, we can easily identify the edges those have more weights, which represent the users who have common interests as well their relationship between them. Thus, lead to effective and credible community. After analysis, we have got community as subgraph $RG$ marked on Graph G as shown on Figure 4. Some problems have been stated regarding detection of communities from social networks. Different mining algorithms use global criterion to identify a subgraph as a community. But fixed criterion tends to clusters having much vertices or users. Many of those are not really belongs to the community. Previous researches on community detection have limitation on overlapping communities.

V. DISCUSSION

Different analysis does not consider both edge contents and vertex contents, which are vital on different community feature extraction. There were also limitations regarding the differentiation of stable topics and temporal topics. As we know from the Bin Cui's [10] research that temporal topics can be the best measure of connectivity by calculating the burst features. Those burst features are identified from given context keys. Our detected community is shown as subgraph on Figure 4 and the result also can be found on the matrix on Figure 5.

VI. CONCLUSION

In this paper, we have proposed a approach that not dependent one single criterion. We have proposed Interlinked Spatial Clustering Model (ILSCM), which can detect community based on both edge and vertex contents.The model also considers the temporal topics which for identifying betweenness. Our approach is to count on those links where there is interlink or edge between the users or nodes by like, comment, share, page, group, location, tags etc which refers better connectivity. Our model will perform a search based on context key provided. Based on the context key, set of burst words and related contents will be extracted from each edges and their associated nodes. We have established the count of bursts as weights on every edges that will identify the close betweenness. Thus a threshold will form the subgraph which can be detected as community. According to the study of the other models, our model will be more effective on community detection because it will consider the interlinked contents and sort out stable and temporal topics, working as a filter to large graph which lead to proper community detection.

## References

[1] H. Yin, B. Cui, H. Lu, Y. Huang, and J. Yao, "A unified model for stable and temporal topic detection from social media data," Data Engineering(ICDE), 2013 IEEE 29th International Conference on, pp. 661–672,2013.

[2] Z. Hu, C. Wang, J. Yao, E. Xing, H. Yin, and B. Cui, "Community Specific Temporal Topic Discovery from Social Media," no. Journal Article, 2013.

[3] D. Goldstein and M. Langberg, "The Dense k Subgraph problem," Algorithmica, vol. 29, no. 3, pp. 410–421, 2009.

[4] M. Delgado and K. Barton, "Murals in latino communities: Social indicators of community strengths," Social work, vol. 43, no. 4, pp. 346–356, 1998.

[5] S. Gregory, "An algorithm to find overlapping community structure in networks," Knowledge Discovery in Databases PKDD 2007, vol. 4702, pp. 91–102, 2007.

[6] M. E. J. Newman, "Analysis of weighted networks," Phys. Rev. E, vol. 70, p. 056131, Nov 2004.

[7] G. J. Qi, C. C. Aggarwal, and T. Huang, "Community detection with edge content in social media networks," Proceedings – International Conference on Data Engineering, pp. 534–545, 2012.

[8] Z. Hu, J. Yao, and B. Cui, "User Group Oriented Dynamic Exploration," pp. 66–72, 2013.

[9] J. Yang, "Defining and Evaluating Network Communities based on Ground-truth,"

[10] W. Cui, Y. Xiao, H. Wang, Y. Lu, and W. Wang, "Online search of overlapping communities," Proceedings of the 2013 international conference on Management of data - SIGMOD '13, p. 277, 2013.

[11] B. J. Yang and J. Leskovec, "Explain Core Periphery Organization of Networks," vol. 102, no. 12, 2014.

[12] Z. Hu, J. Yao, B. Cui, and E. P. Xing, "Community Level Diffusion Extraction," Sigmod'15, pp. 1555–1569, 2015.

[13] J. Yang, "Overlapping Community Detection at Scale : A Nonnegative Matrix Factorization Approach," 2013.

[14] K. Avrachenkov, Algorithms and models for the web-graph 6th international workshop, WAW 2009, Barcelona, Spain, February 12-13, 2009 : proceedings. Berlin: Springer, 2009.

[15] D. Hallac, J. Leskovec, and S. Boyd, "Network Lasso : Clustering and Optimization in Large Graphs,"

[16] J. Yang, "Community-Affiliation Graph Model for Overlapping Network Community Detection," vol. 1.

[17] J. Yang and J. Mcauley, "Community Detection in Networks with Node Attributes,"

[18] M. Kim, "Modeling Social Networks with Node Attributes using the Multiplicative Attribute Graph Model,"

[19] A. B. Lawson and D. G. Denison, Spatial cluster modelling. CRC press, 2002.

[20] X. Wang, C. Zhai, X. Hu, and R. Sproat, "Mining correlated bursty topic patterns from coordinated text streams," in Proceedings of the 13th ACM SIGKDD International Conference on Knowledge Discovery and Data Mining, KDD '07, (New York, NY, USA), pp. 784–793, ACM, 2007.

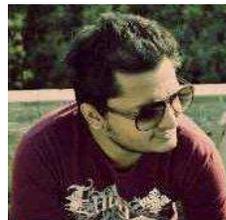

**Victor Stany Rozario** completed B.Sc. in Computer Science & Engineering and M.Sc. in Computer Science from American International University-Bangladesh, Dhaka, Bangladesh. His current research interest includes Data Science, Data Mining, Intelligent Systems, and Human Computer Interaction.

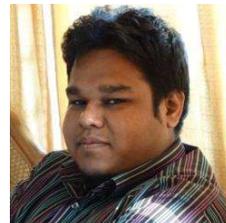

**AZM Ehtesham Chowdhury** completed B.Sc. in Computer Science & Engineering and M.Sc. in Computer Science from American International University-Bangladesh, Dhaka, Bangladesh. His current research interest includes Data Science, Data Analytics, Software Engineering and Mining, Intelligent Systems. Computer Vision, Pattern Recognition , Human Centred Technology (HCT), and Human Centred Design (HCD).

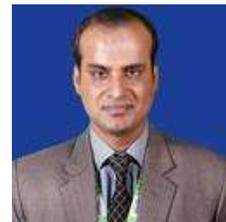

**Muhammad Sarawar Jahan Morshed** completed his B.Sc. in Computer Science & Engineering from Khulna University, Bangladesh and M.Sc. in Computer Science and Engineering from KTH-Royal Institute of Technology, Sweden. Currently he is working as a Assistant Professor in the Department of Computer Science, AIUB. His research interests include Communication Security, Real Time Data Analytic, Evolutionary Computing.